\documentclass[prc,aps,superscriptaddress,showpacs,nofootinbib,twocolumn]{revtex4}
\usepackage{amssymb}
\usepackage[dvips]{graphicx}
\usepackage[english]{babel}
\usepackage{indentfirst}
\usepackage{amsxtra}
\usepackage{amsmath}
\usepackage{supertabular}
\usepackage{multirow}
\usepackage[mathcal]{eucal}

\usepackage{soul}
\usepackage[usenames]{color}
\usepackage{ulem}

\begin{document}
\title {\textbf{Soft breathing modes in neutron--rich nuclei with the subtracted second random--phase approximation}}
\author{D. Gambacurta}
\affiliation{Extreme Light Infrastructure - Nuclear Physics (ELI-NP), Horia Hulubei National Institute for Physics and Nuclear
Engineering, 30 Reactorului Street, RO-077125 M˘agurele, Jud. Ilfov, Romania}
\author{M. Grasso}
\affiliation{Institut de Physique Nucl\'eaire, CNRS-IN2P3, Universit\'e Paris-Sud,
Universit\'e Paris-Saclay, 91406 Orsay, France}
\author{O. Sorlin}
\affiliation{Grand Acc\'el\'erateur National d'Ions Lourds (GANIL), CEA/DSM-CNRS/IN2P3, BP 55027, F-14076, Caen Cedex 5, France}

\begin{abstract}
We analyze the isoscalar response related to breathing modes with particular attention being paid to low-lying excitations in neutron--rich nuclei. We use the subtracted second random--phase approximation (SSRPA) to describe microscopically the response. 
By increasing the neutron excess, we study the evolution of the response in Ca isotopes going from $^{40}$Ca to $^{48}$Ca and to $^{60}$Ca as well as in 
$N=20$ isotones going from $^{40}$Ca to $^{36}$S and to $^{34}$Si. Finally, the case of $^{68}$Ni is investigated. We predict soft monopole modes in 
neutron--rich nuclei which are driven by neutron excitations. At variance with dipole pygmy modes, these neutron excitations are not only strongly dominant at the surface of the nucleus but over its entire volume. The effect of the mixing with two particle-two hole configurations induced by the SSRPA model is analyzed. The properties of such soft neutron modes are 
investigated in terms of their excitation energies, transition densities and wave--function components. Their collectivity is also discussed as a function of the isospin asymmetry and of the mass of the nucleus. The link between such low--energy compression modes and a compressibility modulus introduced for neutron--rich infinite matter is finally studied.
\end{abstract}

\maketitle
% {\color{red} prova }
\section{Introduction}
\label{intro}
Excitation spectra in neutron--rich nuclei are tightly connected to their strong isospin asymmetry.  Exotic excitation modes may occur which are related to the existence of a neutron excess. The most well known examples are the so--called pygmy resonances. 
Pygmy dipole resonances have been widely studied experimentally (see, for example, Refs. 
\cite{eriksen,toft,massa,sch,ozel,krum,endres,pellegri,martorana} and Refs. \cite{savran,bracco} for review articles) and their complex nature has been extensively discussed and interpreted in terms of oscillations of a neutron skin against a core, mixing of isoscalar and isovector nature, coupling with a toroidal motion, single--particle versus collective behavior, deformation effects (see for instance some recent publications \cite{dietz,sun,neste,gamba,wang,kimura,degre,egorova,lanza,tsone,paar,burrello}). Experimental evidence on the existence of low--energy excitations related to a neutron excess was also reported in Refs. \cite{PQR1,PQR2} for the quadrupole channel.

On the other side, soft monopole excitations have been much less analyzed in neutron--rich nuclei so far. 
They were predicted theoretically for  neutron--rich Ca \cite{capelli}, Ni \cite{capelli,khan2011,hamamoto}, Pb and Sn \cite{khan2013} isotopes. A few attempts to 
measure them in exotic nuclei have been performed but a clear experimental signature of their existence is still missing. In the last decade,  
measurements were done on neutron--rich Ni isotopes, using active targets as detectors. This type of experimental studies was carried out for $^{56}$Ni with deuterons as probes \cite{charlotte} and for $^{68}$Ni using $\alpha$ and deuteron scattering \cite{marine}. 
In particular, based on theoretical predictions, $^{68}$Ni was expected  to be a good candidate for presenting a soft breathing mode. 
However, several limitations of the experimental setup could not allow for a clear observation of soft monopole excitations in the measurement described in Ref. \cite{marine} for $^{68}$Ni and the authors pointed out the necessity of upgraded active-target setups. On the other side, most of the available theoretical calculations \cite{capelli,khan2011,khan2013} are based on the mean--field approach. It may then be interesting to have new insights into these excitation modes performing a theoretical investigation that encompasses beyond--mean--field effects and that may, for this reason, provide a clearer and more complete analysis of their nature. This is done in the present article by using the subtracted second random--phase approximation (SSRPA) introduced in Refs. \cite{gamba2015,epja}.     
Two particle-two hole (2p2h) configurations are coupled with one particle--one hole (1p1h) elementary excitations and a subtraction of the self--energy is carried out to remove the double counting of correlations (if traditional effective functionals are used for such beyond--mean--field models), the  instabilities related to the violation of the Thouless theorem \cite{papa}, and the possible ultraviolet divergences (in cases where zero--range effective interactions are used), as suggested in Ref. \cite{tse}. Details on such a subtraction procedure 
applied to the second random--phase--approximation model  
may be found in Ref. \cite{gamba2015}.

% We analyze soft monopole modes using the Skyrme parametrization SGII \cite{sgii,sgii1}. The residual interaction is fully taken into account in the SSRPA model and includes all the rearrangement tems \cite{JPG}. A numerical cutoff of 60 MeV is used to truncate on the excitation energies of 2p2h configurations (we recall however that SSRPA results are not affected by the choice of this cutoff value \cite{gamba2015}). The same cutoff value is used for computing the corrective terms in the subtraction procedure and a diagonal approximation is employed in such a computation for the 2p2h sector of the matrix (that has to be inverted to calculate the corrective term).   
The paper is organized as follows. In Sec. \ref{ND} we provide all the numerical details for the calculations carried out in the present work. In Sec. \ref{softca} we  consider the case of the $N=20$ $N=Z$ nucleus $^{40}$Ca to check whether a low--lying monopole excitation is predicted for this nucleus (having no neutron excess) and which is its nature. In the same section, more neutron--rich Ca isotopes are analyzed, $^{48}$Ca ($\delta=(N-Z)/A=0.17$) and $^{60}$Ca ($\delta=0.33$), and the evolution of the physical nature of the predicted low--lying strength is investigated going from $N=20$ to $N=28$. The strongly isospin asymmetric case of $^{60}$Ca is studied as an illustration to check how these excitations, and in particular their collectivity, further evolve with the neutron excess. Section  \ref{soft20} focuses on $N=20$ isotones and the neutron--rich nuclei 
 $^{34}$Si and  $^{36}$S ($\delta=0.18$ and 0.11, respectively) are analyzed, to be compared with the $N=Z$ nucleus $^{40}$Ca. Section \ref{nickel} is finally dedicated to $^{68}$Ni ($\delta=0.18$), where the isospin asymmetry is comparable to that of $^{48}$Ca and $^{34}$Si, but the number of nucleons is higher. This is done to identify a possible dependence of the collectivity of low--lying excitations 
on the mass of the nucleus. 
An enhanced collectivity in heavier nuclei would allow for an easier experimental observation in those cases. 
Finally, Sec. \ref{compre} describes a link between the energies of such soft compression modes and a compressibility modulus introduced for neutron--rich infinite matter.
Conclusions are drawn in Sec. \ref{conclu}. 

\begin{figure}
\includegraphics[scale=0.37]{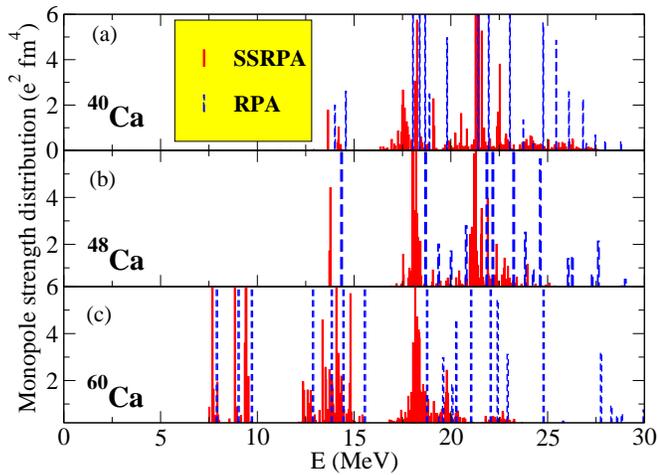}
\caption{(a) Monopole strength distribution computed with RPA (dashed blue bars) and SSRPA (full red bars) for $^{40}$Ca; (b) Same as in (a) but for $^{48}$Ca; (c) Same as in (a) but for $^{60}$Ca.  }
\label{soft40}
\end{figure}

\section{Numerical details}
\label{ND}

 The formalism and the details of the SSRPA model can be found in Ref. \cite{gamba2015}. 
 The calculation scheme is fully self-consistent, which means that the residual interaction is  consistently employed in the SSRPA model with respect to the ground--state calculations carried out with the mean--field Hartree-Fock model. In addition, the residual interaction includes all the rearrangement tems \cite{JPG}. The Skyrme parametrization SGII \cite{sgii,sgii1} is used.  
 
 A cutoff of 80 MeV is chosen for building the 1p1h configurations, ensuring a full preservation of the isoscalar and isovector energy--weighted sum rules (EWSRs) at the level of the random--phase approximation (RPA). Deviations of less than 1\% are found.  
 A numerical cutoff of 60 MeV is used to truncate on the excitation energies of 2p2h configurations (we recall, however, that the SSRPA results are not affected by this cutoff and are stable with respect to the choice of its value \cite{gamba2015}).

 The same cutoff value on 2p2h configurations is used  both for the construction of the matrix to be 
diagonalized and for the evaluation of the corrective terms induced by the  subtraction procedure \cite{gamba2015}.
The evaluation of these corrective terms is performed by using a diagonal approximation. We have shown that this approximation does not affect the results \cite{gamba2015}. The interaction between 2p2h configurations is instead fully taken into account in the 2p2h block matrix of the SSRPA eigenvalue problem.  The SSRPA model does not account explicitly for pairing correlations. Work to treat superfulid systems is presently in progress. For this reason, we have chosen to limit our analysis to systems having both proton and neutron shell or subshell closures, with therefore a limited (or negligible) amount or pair correlations as compared to mid-shell nuclei.

\section{Ca isotopes with neutron excess equal to 0, 0.17, and 0.33}
\label{softca}
\begin{figure}
\includegraphics[scale=0.32]{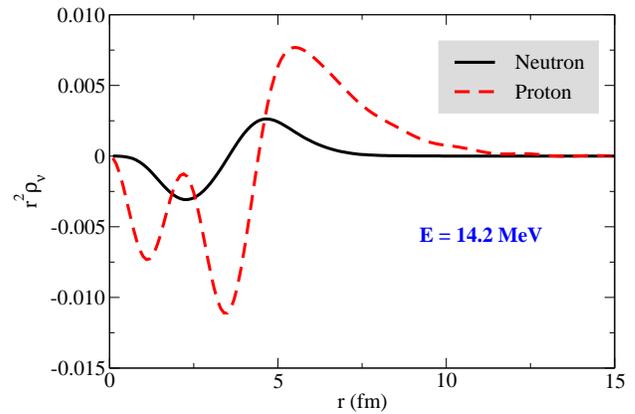}
\caption{Neutron and proton transition densities multiplied by $r^2$ (in units of fm$^{-1}$) for $^{40}$Ca  associated with the SSRPA energy peak located at 14.2  MeV. }
\label{td40}
\end{figure}

\begin{figure}
\includegraphics[scale=0.32]{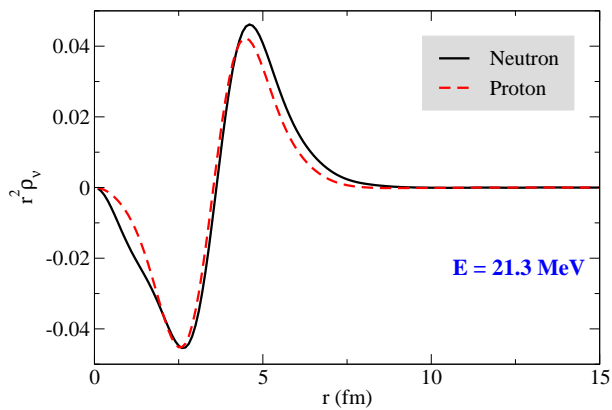}
\caption{Same as in Fig. \ref{td40} but for the peak located at 21.3  MeV. }
\label{GMR40}
\end{figure}

A first investigation is done on $^{40}$Ca (having no neutron excess) and the corresponding SSRPA monopole strength distribution is plotted in Fig. 
\ref{soft40}(a). In addition to the region of the isoscalar giant monopole resonance, which may be easily recognized in the figure, one observes the presence of some strength 
for this nucleus in the energy region around 14 MeV. For comparison, also the RPA strength distribution is shown to indicate the main effects provided by the beyond--mean--field SSRPA model, compared to the mean--field--based RPA predictions. 
A low--lying excitation is found also with the RPA model, the main difference being a shift to lower energies  and a stronger fragmentation produced by SSRPA owing to the coupling with 2p2h configurations. To analyze the nature of these excitation modes, the SSRPA neutron and proton transition densities multiplied by $r^2$ and  associated with the energy peak located at 14.2  MeV are shown in Fig. \ref{td40}. The dominant proton contribution can be clearly seen. It was in particular found that 
such an excitation is mainly driven by the proton 1p1h configuration $[\pi 3s_{1/2}, \pi 2s_{1/2}]^{J=0}$. The RPA prediction is analougous, apart from the fact that (i) an energy shift exists between the two spectra, (ii) only 1p1h configurations compose this excitation in RPA, and (iii) there is a higher fragmentation of the strength in the SSRPA spectrum. In the SSRPA case, the mixing with higher--order  configurations leads to a strong contribution coming from the 2p2h sector. The 2p2h contribution to the norm of the state is indeed close to 60\%  and the main configurations are shown in Table \ref{Tab:Ca40}. 
The percentage of the EWSR computed in SSRPA up to 15 MeV is 2.13\%. 

To make clear the difference between this low--lying part of the spectrum and the isoscalar giant monopole resonance, SSRPA neutron and proton transition densities multiplied by $r^2$ are plotted in Fig. \ref{GMR40} for the peak located at 21.3 MeV. One can observe that, in this case, neutrons and protons participate together and coherently to the excitation mode.

\begin{widetext}

\begin {table} 
\begin{center}
\begin{tabular}{ccc}
                     \hline
\hline
   $^{40}$Ca    &  1p1h & 2p2h  \\
          &    40 \%  & 60 \%  \\
\hline
  & $[\pi 3s_{1/2}, \pi 2s_{1/2}]^{J=0}$ & $[[\pi5p_{3/2},\pi2f_{7/2}]^{J_P=2}[\pi1d_{5/2},\pi1d_{3/2}]^{J_H=2}\big]^{J=0}$       \\
 & & $[[\pi 4s_{1/2},\nu 3d_{5/2}]^{J_P=3}[\pi 1d_{5/2},\nu 1d_{3/2}]^{J_H=3}\big]^{J=0}$   \\
 & & $[[\pi 4p_{3/2},\nu 3f_{5/2}]^{J_P=4}[\pi 1d_{5/2},\nu 1d_{3/2}]^{J_H=4}\big]^{J=0}$   \\
 & & $[[\pi 3f_{7/2},\nu 1f_{7/2}]^{J_P=3}[\pi 1d_{5/2},\nu 1d_{3/2}]^{J_H=3}\big]^{J=0}$   \\
 & & $[[\pi 4s_{1/2},\pi 5d_{5/2}]^{J_P=2}[\pi 1d_{3/2},\pi 1d_{3/2}]^{J_H=2}\big]^{J=0}$   \\
 & & $[[\pi 2p_{1/2},\nu 2d_{3/2}]^{J_P=1}[\pi 1d_{3/2},\nu 1p_{3/2}]^{J_H=1}\big]^{J=0}$ \\
 & & $[[\pi 4p_{3/2},\nu 1f_{7/2}]^{J_P=2}[\pi 1d_{3/2},\nu 1d_{5/2}]^{J_H=2}\big]^{J=0}$  \\
 & & $[[\pi 3f_{7/2},\nu 1f_{7/2}]^{J_P=3}[\pi 1d_{3/2},\nu 1d_{5/2}]^{J_H=3}\big]^{J=0}$ \\
 & & $[[\pi 2d_{3/2},\nu 6s_{1/2}]^{J_P=2}[\pi 1d_{3/2},\nu 2s_{1/2}]^{J_H=2}\big]^{J=0}$ \\
 & & $[[\nu 4s_{1/2},\nu 4d_{5/2}]^{J_P=2}[\nu 2s_{1/2},\nu 1d_{3/2}]^{J_H=2}\big]^{J=0}$    \\ 
\hline
\end{tabular}
\end{center}
\caption{Composition of the peak located at 14.2 MeV for $^{40}$Ca. The percentages of the total contributions coming from 1p1h and 2p2h configurations are indicated and the dominant configurations are listed.  }
\label{Tab:Ca40}
\end {table} 
\end{widetext}

We then illustrate the case of a neutron excess equal to 0.17 with the Ca isotope $^{48}$Ca. Figure \ref{soft40}(b) shows the strength distributions in the RPA and SSRPA cases. Also this time the effect of SSRPA is to produce a shift of the strength distribution to lower energies and to induce more fragmentation in the spectrum, compared to the RPA case. 

The SSRPA peak located at 13.8 MeV is mainly composed by the  1p1h configuration $[\nu 2f_{7/2},\nu 1f_{7/2}]^{J=0}$ and by the 2p2h configuration $[ \nu 2f_{7/2}, \nu 4f_{7/2}]^{J_P}[ \nu 1f_{7/2}, \nu1f_{7/2}]^{J_H}]^{J=0}$, with $J_P=J_H=$ 0, 2, 4, the dominant component being the one with $J_P=J_H=4$ .  
The other 2p2h configurations lead to negligible contributions to the peak and for this reason are not reported on a table.

\begin{figure}
\includegraphics[scale=0.32]{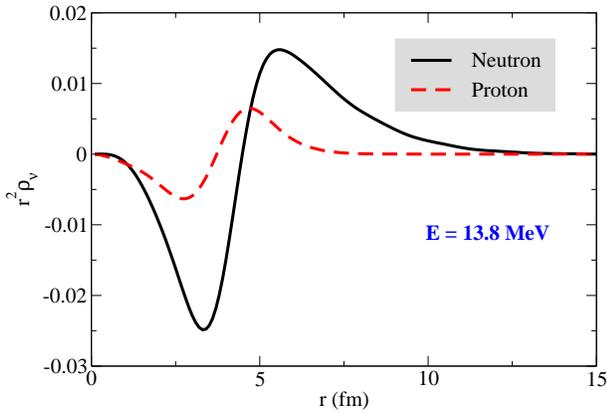}
\caption{Same as in Fig. \ref{td40} but for $^{48}$Ca. The peak for which the transition densities are calculated is located at 13.8 MeV.  }
\label{td48}
\end{figure}

SSRPA neutron and proton transition densities are shown in Fig. \ref{td48} for this peak. It may be seen that the physical nature of this excitation mode has strongly evolved compared to the case of $^{40}$Ca. Now this excitation is mainly driven by neutrons. 
The percentage of the EWSR computed up to 15 MeV is 2.5\%, slightly larger than for $^{40}$Ca.
The transition probability of the state is mainly composed by the most important 1p1h contribution but several other 1p1h configurations (mainly neutron configurations) contribute to the total $B(E0)$. Consequently, this state is slightly more collective (but still weakly collective) than the proton state in $^{40}$Ca.
As could be expected, by increasing the neutron excess from $N=20$ to $N=28$, the low--lying excited mode has acquired a dominant  neutronic nature even if, however, a small proton contribution ia also predicted. As indicated by the transition densities, the dominant neutron contribution is not only localized at the surface of the nucleus but extends over the entire volume of the system. 

\begin{figure}
\includegraphics[scale=0.32]{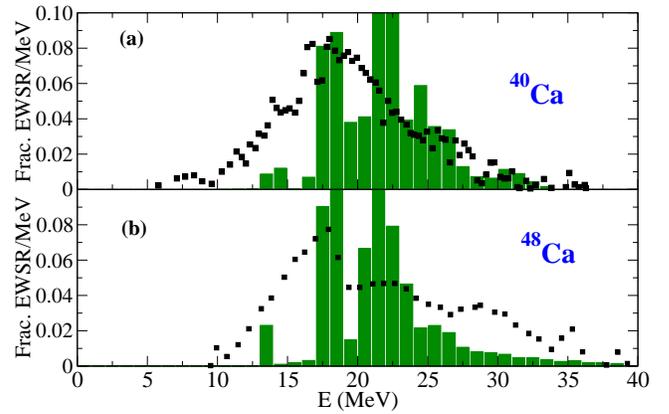}
\caption{(a) Black squares: experimental results of Ref. \cite{lui}; green bars: SSRPA predictions for $^{40}$Ca; (b) Same as in (a) but for $^{48}$Ca.  }
\label{gmr40}
\end{figure}

Since the isoscalar giant monopole strength distributions were measured for both $^{40}$Ca and $^{48}$Ca \cite{lui}, we show in Fig. \ref{gmr40} the comparison between the SSRPA predictions  and the experimental results. The fractions of EWSR/MeV are reported in Figs. \ref{gmr40}(a) and \ref{gmr40}(b) for $^{40}$Ca and $^{48}$Ca, respectively. The SSRPA results indicate how the SSRPA model significantly improves the description of the fragmentation of the strength compared to the RPA case (the RPA discrete spectra are plotted in Fig. \ref{soft40}(a) and (b) for the two nuclei). Figure 10 of Ref. \cite{lui} shows the comparison between the experimental distributions for $^{40}$Ca and $^{48}$Ca and the corresponding folded (thus, with an artificial width) RPA distributions with several Skyrme parametrizations. One may observe from that figure that the RPA model systematically predicts strength distributions which are shifted at higher energies compared to the experimental ones. We have calculated the centroid energies produced by the parametrization SGII in the RPA (SSRPA) model and obtained 21.3 (20.7) and 20.7 (20.4) MeV for $^{40}$Ca and $^{48}$Ca, respectively. The centroids are computed using $\sqrt{m_1/m_{-1}}$, where $m_1$ and 
$m_{-1}$ are the energy--weighted and the inverse energy--weighted moments of the strength distribution, respectively. The experimental values for the centroids are 18.3 and 19.0 MeV for  $^{40}$Ca and $^{48}$Ca, respectively. 
These results indicate that the SSRPA predictions lead to a better agreement (compared to RPA) with these known experimental values.  

\begin{figure}
\includegraphics[scale=0.32]{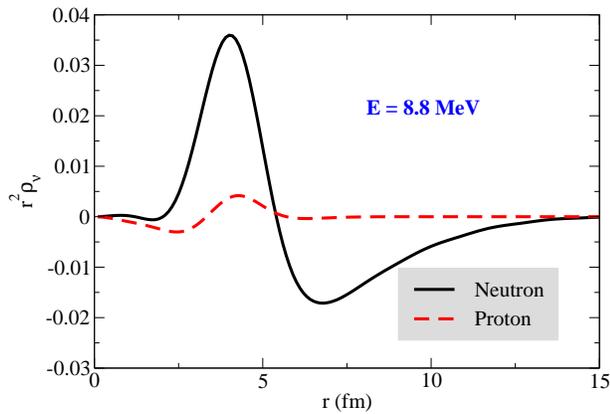}
\caption{Same as in Fig. \ref{td40} but for $^{60}$Ca. The peak for which the transition densities are calculated is located at 8.8 MeV.  }
\label{td60}
\end{figure}

Finally, the extreme case of $^{60}$Ca is studied. The discovery of this isotope at the RIKEN Nishina Center was recently reported in Ref. \cite{tarasov}.  The SSRPA and RPA strength distributions are plotted in Fig. \ref{soft40}(c). One observes 
the same effects as for the previous two cases in the comparison between the RPA and SSRPA spectra. 
For this extreme case, the strength distribution is localized in three main windows: from 5 to 11 MeV, from 11 to 16 MeV and, finally, at energies larger than 16 MeV (giant monopole resonance). 
The first region of the strength distribution is not predicted for the less neutron--rich nucleus $^{48}$Ca. 
We investigate the nature of the excitations in the first two energy regions by analyzing some corresponding transition densities.  The peak located at 8.8  MeV
is chosen to represent the first energy window and the corresponding transition densities are plotted in Fig. \ref{td60}. A dominant neutron contribution is found also in this case, as for $^{48}$Ca (the effect is more pronounced). This state is mainly composed (70\%) by the 1p1h configuration $[\nu 2f_{5/2},\nu 1f_{5/2}]^{J=0}$. The remaining 30\% is provided by highly fragmented 2p2h neutron configurations, for which it is not easy to  identify some dominant configurations.  The percentage of the EWSR computed up to 11 MeV is equal to 5.13\%, which indicates an enhanced collectivity compared to the low--lying part of the spectrum in $^{48}$Ca.   

\begin{figure}
\includegraphics[scale=0.32]{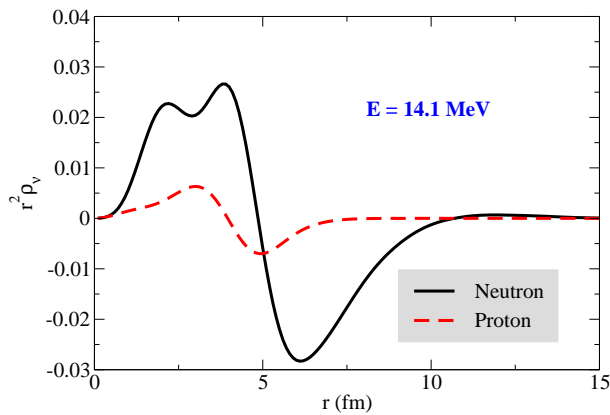}
\caption{Same as in Fig. \ref{td40} but for $^{60}$Ca. The peak for which the transition densities are calculated is located at 14.1 MeV.}
\label{td60b}
\end{figure}

In the energy region from 11 to 16 MeV, the most collective state is located at 14.1 MeV. It is a state with a dominant 2p2h nature (86\%) with two strong 1p1h configurations, $[\nu 4p_{3/2}, \nu 2p_{3/2}]^{J=0}$ and $[ \nu 3f_{5/2}, \nu 1f_{5/2}]^{J=0}$. Also in this case the 2p2h components are highly fragmented.  The corresponding transition densities are plotted in Fig. \ref{td60b}.
The percentage of the EWSR computed up to 16 MeV is equal to 26.81\%. 
This means that more than 20\% of the contribution to the EWSR is produced between 11 and 16 MeV whereas only 5\% comes from lower energies. 
This result indicates that the states located in the second energy are much more collective. The nature of these low--lying excitation modes is the same in the two nuclei $^{48}$Ca and $^{60}$Ca (neutron excitation extending over the entire volume of the nucleus) but the collectivity is strongly increased in $^{60}$Ca. Comparing the states located around 14 MeV in $^{48}$Ca and in $^{60}$Ca, one observes that the excitations have changed from mostly single--particle excitations with low collectivity to states with a dominant 2p2h nature, where numerous configurations are mixed to build the excitation modes and the collectivity is highly enhanced.  

For the other systems that will be studied in this work, the comparison with the RPA spectrum will not be illustrated because the found results have always the same trend: the excitation modes are  found also in RPA, but the SSRPA model leads to a shift to lower energies and to spreading effects produced by the mixing with 2p2h configurations.  

\section{$N=20$ isotones. Soft modes in $^{34}$Si and $^{36}$S.}
\label{soft20}

After having analyzed the evolution of low--lying monopole modes in Ca isotopes going from $N=20$ to $N=40$, we present in this section a similar analysis done this time with $N=20$ isotones, going from $^{40}$Ca to the neutron--rich nuclei $^{36}$S and $^{34}$Si. 

\begin{figure}
\includegraphics[scale=0.32]{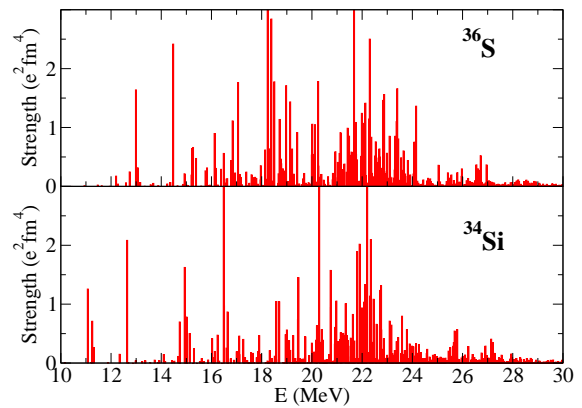}
\caption{Monopole isoscalar strength distributions calculated for the nuclei $^{36}$S (a) and $^{34}$Si (b).  }
\label{soft3436}
\end{figure}

Figure \ref{soft3436} shows the isoscalar monopole strength distributions in $^{36}$S and $^{34}$Si. 
One observes the presence of some strength above 10 MeV for both nuclei. For $^{36}$S there is a first peak placed at around 13 MeV whereas for $^{34}$Si the first peak is located about 2 MeV downwards. The analysis of the nature of these excitations tells us that there is a non negligible  mixing between 1p1h and 2p2h excitations and that the 1p1h contribution is driven in both nuclei by the neutron single--particle configuration $[\nu 2d_{3/2}, \nu 1d_{3/2}]^{J=0}$. The difference of $\sim$ 2 MeV in the excitation energies of the first peaks for the two nuclei is essentially due to the difference in the energies of the dominant neutron single--particle configuration. For $^{34}$S ($^{36}$S) the 1p1h contribution to the peak located at 11.07 (12.99) MeV is 54 (52) \%. The rest of the contribution is given by the mixing of several 2p2h configurations. The most important configurations are listed in Table \ref{tab1}. 
For $^{34}$Si the 1p1h configuration 
$[\nu 2d_{3/2}, \nu 1d_{3/2}]^{J=0}$ 
contributes to 53 \% of the total composition of this excitation whereas the listed 2p2h configurations 
contribute together to 31 \%. For $^{36}$S the 1p1h configuration 
$[\nu 2d_{3/2}, \nu 1d_{3/2}]^{J=0}$ 
contributes to 51 \% of the total composition and the 
listed 2p2h configurations contribute together to 29\%.

\begin{widetext}

\begin {table} 
\begin{center}
\begin{tabular}{ccc}
                     \hline
\hline
   $^{34}$Si    &  1p1h & 2p2h  \\
          &    54 \%  & 46 \%  \\
\hline
  & $[\nu 2d_{3/2}, \nu 1d_{3/2}]^{J=0}$ & $[[\pi 3p_{1/2}, \nu 3f_{7/2}]^{J_p=3}[\pi 1d_{5/2}, \nu 2s_{1/2}]^{J_h=3}]^{J=0}$       \\
 & & $[[\pi 4p_{1/2}, \nu 1f_{5/2}]^{J_p=2}[\pi 1d_{5/2}, \nu 2s_{1/2}]^{J_h=2}]^{J=0}$   \\
 & & $[[\pi 4p_{1/2}, \nu 1f_{5/2}]^{J_p=3}[\pi 1d_{5/2}, \nu 2s_{1/2}]^{J_h=3}]^{J=0}$   \\
 & & $[[\pi 6s_{1/2}, \nu 2d_{3/2}]^{J_p=2}[\pi 1d_{5/2}, \nu 1d_{3/2}]^{J_h=2}]^{J=0}$   \\
 & & $[[\pi 6s_{1/2}, \nu 2d_{5/2}]^{J_p=2}[\pi 1d_{5/2}, \nu 1d_{3/2}]^{J_h=2}]^{J=0}$   \\
 & & $[[\pi 3d_{3/2}, \nu 3s_{1/2}]^{J_p=2}[\pi 1d_{5/2}, \nu 1d_{3/2}]^{J_h=2}]^{J=0}$ \\
 & & $[[\pi 3d_{3/2}, \nu 2d_{5/2}]^{J_p=1}[\pi 1d_{5/2}, \nu 1d_{3/2}]^{J_h=1}]^{J=0}$  \\
 & & $[[\pi 3d_{3/2}, \nu 2d_{5/2}]^{J_p=2}[\pi 1d_{5/2}, \nu 1d_{3/2}]^{J_h=2}]^{J=0}$ \\
 & & $[[\pi 3d_{3/2}, \nu 2d_{5/2}]^{J_p=3}[\pi 1d_{5/2}, \nu 1d_{3/2}]^{J_h=3}]^{J=0}$ \\
 & & $[[\nu 3d_{3/2}, \nu 2d_{5/2}]^{J_p=2}[\nu 1d_{3/2}, \nu 1d_{3/2}]^{J_h=2}]^{J=0}$    \\ 
\hline
   $^{36}$S    &  1p1h & 2p2h   \\
          &    52 \%  & 48 \%  \\
\hline
  & $[\nu 2d_{3/2}, \nu 1d_{3/2}]^{J=0}$ &    
  $[[\pi 3d_{3/2}, \nu 4d_{3/2}]^{J_p=2}[\pi 1d_{5/2}, \nu 1d_{3/2}]^{J_h=2}]^{J=0}$ \\
 & & $[[\pi 4d_{3/2}, \nu 4s_{1/2}]^{J_p=2}[\pi 2s_{1/2}, \nu 1d_{3/2}]^{J_h=2}]^{J=0}$  \\
 & & $[[\pi 4d_{3/2}, \nu 5s_{1/2}]^{J_p=1}[\pi 2s_{1/2}, \nu 1d_{3/2}]^{J_h=1}]^{J=0}$  \\
 & & $[[\pi 4d_{3/2}, \nu 4d_{3/2}]^{J_p=2}[\pi 2s_{1/2}, \nu 1d_{3/2}]^{J_h=2}]^{J=0}$  \\
 & & $[[\pi 4d_{3/2}, \nu 2d_{5/2}]^{J_p=1}[\pi 2s_{1/2}, \nu 1d_{3/2}]^{J_h=1}]^{J=0}$  \\
\hline
\hline
\end{tabular}
\end{center}
\caption{Composition of the peak located at 11.07 (12.99) MeV for $^{34}$Si ($^{36}$S). The percentages of the total contributions coming from 1p1h and 2p2h configurations are indicated and the dominant configurations are listed.  }
\label{tab1}
\end {table} 

\end{widetext}

\begin{figure}
\includegraphics[scale=0.32]{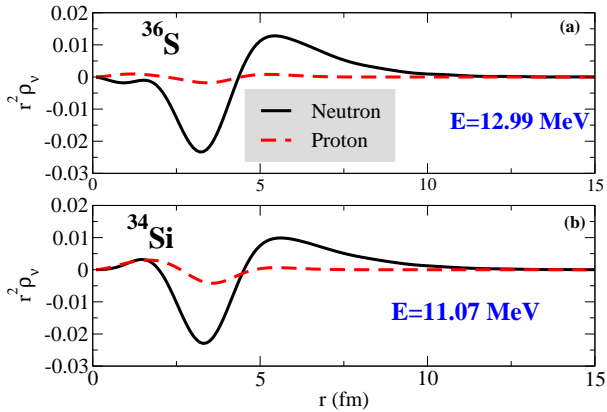}
\caption{Neutron and proton transition densities multiplied by $r^2$ (in units of fm$^{-1}$) for $^{36}$S (a) and $^{34}$Si (b) associated with the energy peak located at 12.99 (11.07) MeV for $^{36}$S ($^{34}$Si). }
\label{trd3436}
\end{figure}

\begin{figure}
\includegraphics[scale=0.32]{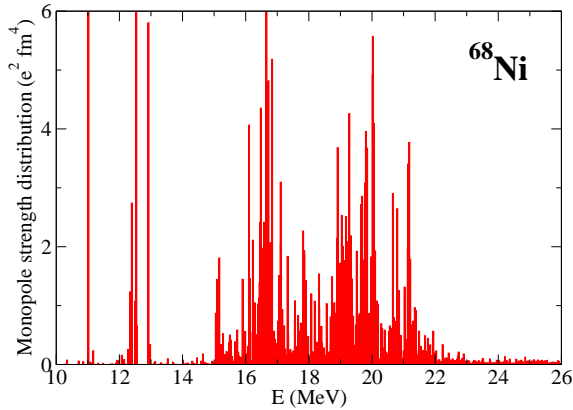}
\caption{Monopole isoscalar strength distribution for $^{68}$Ni. }
\label{soft68}
\end{figure}

Figure \ref{trd3436} shows the neutron and proton transition densities multiplied by $r^2$ for the two nuclei associated 
with the energy peak located at 12.99 (11.07) MeV for $^{36}$S ($^{34}$Si).
The dominant neutron contribution is clearly visible both in the interior and at the surface of the nucleus as for the previous cases $^{48}$Ca and $^{60}$Ca.  The excitation energies are affected by a non negligible mixing between the dominant 1p1h neutron configuration  and several types of 2p2h configurations (the same peaks, composed only by 1p1h configurations, are located 1 MeV upwards in energy in the RPA spectra for the two nuclei). 
The nature of this soft mode is the same for the two $N=20$ isotones under study, the difference between the two nuclei being only the energy location of the first peak and a slightly larger strength found for the most neutron--rich system  $^{34}$Si. To illustrate quantitatively this aspect, we have calculated the percentage of the EWSR in energy regions below the giant resonance for the two nuclei and found that, up to 15 MeV, the percentage of EWSR is equal to 3.15 (4.5) \% for $^{36}$S ($^{34}$Si). 
The low--energy contribution to the EWSR is more important in $^{34}$Si compared to the case of $^{48}$Ca. In spite of the fact that the two nuclei have similar isospin asymmetry, we predict more collectivity in the  case of $^{34}$Si.

\section{The case of $^{68}$Ni}
\label{nickel}

It may be interesting to extend this type of analysis to a heavier neutron--rich nucleus. We choose for this investigation $^{68}$Ni because this nucleus was already indicated as a good candidate for a soft monopole mode with mean--field calculations and because a first experimental study was already carried out for this system, even if definite conclusions could not be drawn from the measurement. By the comparison for example with the lighter nucleus $^{34}$Si (having a similar isospin asymmetry), we wish to check whether the SSRPA model predicts a modification of the nature (and of the collectivity) of the soft mode for a heavier exotic system placed in a different region of the nuclear chart. 
The comparison between $^{34}$Si and $^{68}$Ni is particularly meaningful because these two isotopes present a very similar shell structure. Both of them have for example a double shell closure, of spin--orbit type for protons and of harmonic--oscillator type for neutrons. A possible modification of the nature and of the collectivity of the soft mode in the two nuclei is thus expected to be weakly affected by their respective shell structure and mostly impacted by the mass difference of the two systems. 

\begin{figure}
\includegraphics[scale=0.34]{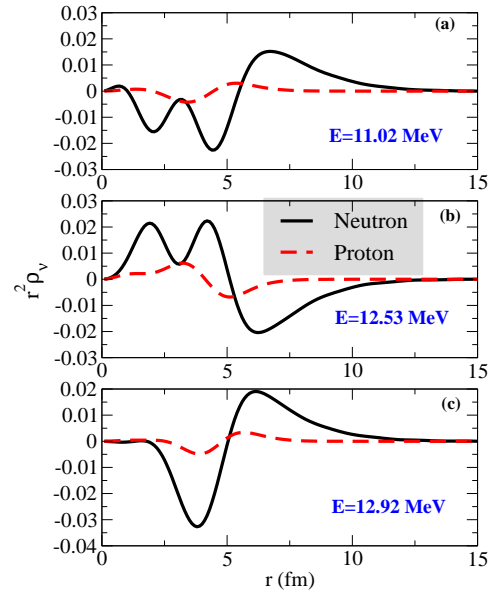}
\caption{ Neutron and proton transition densities multiplied by $r^2$ (in units of fm$^{-1}$) associated with the peaks located at 11.2, 12.53, and 12.92 MeV in the monopole spectrum of $^{68}$Ni.}
\label{td68}
\end{figure}

The response function is plotted in Fig. \ref{soft68} for $^{68}$Ni. 
  One may observe the existence of several peaks located below 14 MeV. The percentage of the EWSR computed below 14 MeV is 5.37 \%. 
Compared to the lighter system $^{34}$Si, the percentage is larger indicating an enhancement of collectivity. This enhancement of collectivity is however much weaker than the one that was predicted going from $^{48}$Ca to  
$^{60}$Ca. The collectivity seems to be more strongly dependent on the neutron excess than on the mass of the system.   
The neutron and proton transition densities corresponding to the three peaks placed at 11.02, 12.53, and 12.92 MeV are plotted in Fig. \ref{td68}.

\begin{figure}
\includegraphics[scale=0.32]{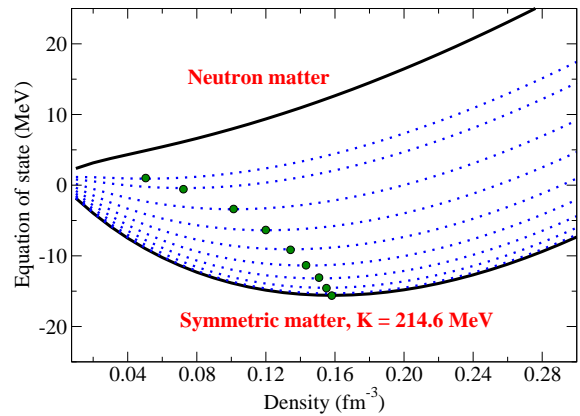}
\caption{Equations of state computed with the parametrization SGII, from symmetric to pure neutron matter. Dotted lines represents the equations of state of asymmetric matter with asymmetry values $X$ (from the bottom to the top) equal to 0.1 ($K_{X}=$ 211.59 MeV), 0.2 ($K_{X}=$ 203.54 MeV), 0.3 ($K_{X}=$ 187.33 MeV), 0.4 ($K_{X}=$ 166.40 MeV), 0.5 ($K_{X}=$ 139.90 MeV), 0.6 ($K_{X}=$ 108.30 MeV), 0.7 ($K_{X}=$ 72.35 MeV), 0.8 ($K_{X}=$ 33.74 MeV), 0.85 ($K_{X}=$ 13.32 MeV). The compressibility value associated with symmetric matter for the parametrization SGII is also indicated in the figure. The equilibrium points (at which the compressibility values $K_X$ are calculated)  are represented by green circles for each equation of state. }
\label{eos}
\end{figure}

\begin{figure}
\includegraphics[scale=0.31]{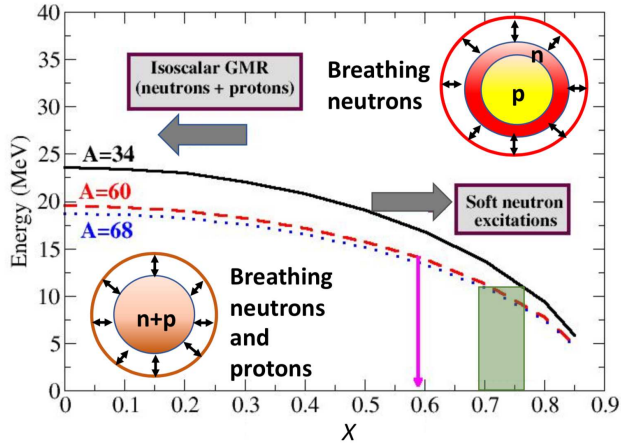}
\caption{Trends provided by Eq. (\ref{eversusk2}) for three values of $A$, 34, 60, and 68. The estimations provided for the centroid of the giant monopole resonance and obtained by using the compressibility value associated with symmetric nuclear matter are the energy values corresponding to $X=$ 0. The green area represents the range of isospin asymmetries predicted for the oscillating system involved in the excitations located at 8.838 ($^{60}$Ca), 11.075 ($^{34}$Si), and 11.021 ($^{68}$Ni) MeV. The magenta arrow indicates the isospin asymmetry associated with the oscillating system involved in the excitation mode located at 14.087 MeV for $^{60}$Ca. }
\label{ekappa}
\end{figure}

For the three peaks there is a dominant neutron 1p1h contribution given by the configurations $[\nu 3p_{1/2},\nu 2p_{1/2}]^{J=0}$, $[\nu 3p_{3/2},\nu 2p_{3/2}]^{J=0}$, and $[\nu 2f_{5/2},\nu 1f_{5/2}]^{J=0}$, respectively, whereas the 2p2h contribution corresponds to 0.18\%,  
0.49 \%, and 0.29 \% of the EWSR for the three cases, respectively. We conclude that the mixing with 2p2h configurations is very weak in this nucleus for the soft monopole excitations under study. 

\section{Neutron compression modes. A link with the equation of state of neutron--rich matter}
\label{compre}

The link existing between the centroid energies of isoscalar giant monopole resonances and the compressibility modulus $K$ of symmetric infinite nuclear matter has been discussed for several decades (see, for instance, Ref. \cite{blaizot} and, for recent reviews, Refs. \cite{roca,colo}). The compressibility modulus for symmetric matter is defined as 
\begin{equation}
K=9 \rho_{0}^2 \Big(\frac{\partial^2 E^{sym}/A}{\partial \rho^2} \Big)_{\rho=\rho_{0}},
\label{comsy}
\end{equation}
where $ E^{sym}/A$ is the equation of state of symmetric matter and $\rho_0$ is the saturation density, equal to 0.16 fm$^{-3}$ for the mean--field equation of state produced for symmetric matter with the parametrization SGII (which corresponds to the accepted empirical value). A more refined and less qualitative link between the centroid energies and a compressibility modulus may be defined by introducing a compressibility modulus for finite nuclei (where several terms appear and the compressibility for symmetric matter corresponds to the volume contribution \cite{blaizot}). We will however remain here on a qualitative level and employ compressibility moduli associated with infinite matter. 

By treating the nucleus as a liquid drop characterized by a compressibility modulus $K$ (the compressibility of symmetric infinite matter) and by linerazing the hydrodynamics equations, one may write a relation between the centroid energy of a giant monopole resonance and the corresponding compressibility modulus,
\begin{equation}
E=\sqrt{\frac{\hbar^2 \pi^2}{15 m}} \sqrt{\frac{K}{\eta_0^2}},
\label{eversusk}
\end{equation}
where $\eta_0$ is the root mean square radius \cite{blaizot}. By using for example $\eta_0=r_0 A^{1/3}$, with $r_0 \sim $ 1 fm, Eq. (\ref{eversusk}) becomes
\begin{equation}
E \sim 5.22 A^{-1/3} \sqrt{K}.
\label{eversusk1}
\end{equation}
The compressibility of symmetric matter is equal to 214.6 MeV for the parametrization SGII used here (and for a mean--field equation of state). By using this value of $K$, Eq. (\ref{eversusk1}) describes qualitatively the evolution of the centroid energies of giant monopole resonances as a function of the mass $A$. 

Now, we have seen in the previous sections that the soft modes which are predicted for neutron--rich nuclei have typically a neutronic nature and, thus, cannot be described as compression modes where neutrons and protons participate together and coherently (that means, as breathing modes where the restoring force can be related to the compressibility modulus of symmetric infinite matter). 
In practice, only neutrons breath in soft modes. 
This also implies that these soft excitations should not be included when estimations for the compressibility modulus of symmetric matter are extracted from measured isoscalar monopole strength distributions. For these estimations, only the giant resonance region should be taken into account. 

We discuss here a different connection, between the energies of these low--energy excitation modes and  a compressibility modulus associated this time with neutron--rich matter. 
This link is intended to be qualitative for two main reasons: (i) it is based on the use of a liquid drop model (as for the link existing between the centroids of giant monopole resonances and $K$); (ii) it is obtained by employing numerical values for the compressibility modulus computed with a mean--field equation of state for infinite matter (whereas our predicted excitation energies are obtained employing a beyond--mean--field model). 

We introduce an isospin--asymmetry parameter $X$. Whereas $\delta$ (introduced in Sec. \ref{intro}) refers to the isospin asymmetry of a given nucleus, $X$, 
equal to $(\rho_n-\rho_p)/\rho$ (where $\rho_n$, $\rho_p$ and $\rho$ are the neutron, proton and total densities, respectively),
 will refer in what follows both to the isospin asymmetry of infinite matter and to the isospin asymmetry of the oscillating system involved in a given excited state. Through the compressibility of asymmetric matter with isospin asymmetry $X$, we will indeed discuss here a link between the isospin  asymmetry of neutron--rich matter and the isospin asymmetry of the oscillating neutron--rich system involved in the low--energy excitation. From the SSRPA microscopic calculations, such an isospin asymmetry may be estimated by introducing the quantities 
\begin{equation}
X_{N}=4 \pi \int |\rho_{\nu}^n| r^2 dr
\label{xnn}
\end{equation}
and
\begin{equation}
X_{P}=4 \pi \int |\rho_{\nu}^p| r^2 dr,
\label{xpp}
\end{equation}
where $\rho_{\nu}^n$ and $\rho_{\nu}^p$ are the microscopic SSRPA neutron and proton transition densities, respectively. 
These quantities estimate the number of neutrons and protons involved in a given excitation $|\nu\rangle$. Since only 1p1h amplitudes enter in the computation of the transition densities, $X_N$ and $X_P$ are a partial estimation taking into account only the 1p1h contribution. The isospin asymmetry $X$ of the oscillating system in a given excited state is related to $X_N$ and $X_P$ by
\begin{equation}
X=\frac{X_N-X_P}{X_N+X_P}.
\label{xtot}
\end{equation}

A compressibility modulus may be introduced for isospin--asymmetric matter,  
\begin{equation}
K_{X}=9 \rho_{eq}^2 \Big(\frac{\partial^2 E^X/A}{\partial \rho^2} \Big)_{\rho=\rho_{eq}},
\label{com1}
\end{equation}
where $E^X/A$ is the equation of state of asymmetric matter and $\rho_{eq}$ is the equilibrium density of such a neutron--rich matter 
 at any isospin asymmetry for which the corresponding equation of state still has an equilibrium point (see Fig. \ref{eos} where the equilibrium points are indicated by green circles for different values of $X$). 

By extending Eq. (\ref{eversusk1}), one may thus write, for a given mass $A$, a relation of the type
\begin{equation}
E(X) \sim 5.22 A^{-1/3} \sqrt{K_{X}},
\label{eversusk2}
\end{equation}
where, when the isospin asymmetry $X$ goes to 0, $K_{X}$ is equal to $K$ and $E(X)$ describes qualitatively the centroid of the giant monopole resonance. Using the values of $K_X$ reported in the legend of Fig. \ref{eos}, one may plot Eq. (\ref{eversusk2}) as a function of the isospin asymmetry for a given value of the mass $A$. This is done in Fig. \ref{ekappa} for the cases where we have predicted, in the corresponding energy window, an EWSR percentage at least equal to 4\%, that is the nuclei for $^{34}$Si, $^{60}$Ca, and $^{68}$Ni.  

We observe that, by increasing the isospin asymmetry of the breathing system in a given nucleus, the excitation energy is expected to decrease compared to the centroid energy of the giant monopole resonance (corresponding to $X=0$). By using the energies of the peaks located at 8.8, 11.07, and 11.02 MeV for $^{60}$Ca, $^{34}$Si, and $^{68}$Ni, respectively, one may deduce from the figure that these soft modes are expected to be characterized by a strong isospin asymmetry in the range 0.69$\div$0.77. 
In particular, we have found $X \sim$ 0.77, 0.76, and 0.69 for the three peaks in  $^{60}$Ca, $^{34}$Si, and $^{68}$Ni, respectively.
These asymmetries are represented by the green region in the figure. For any state with a dominant neutronic nature that we predict in these neutron--rich systems, the reduction of its excitation energy, compared to the isoscalar giant monopole energy (reduction that we predict microscopically), may be qualitatively understood by the trends shown in Fig. \ref{ekappa}. In general, for the lowest states found in the three nuclei, it is expected that the oscillating system is dominantly composed by neutrons with an isospin asymmetry  larger than $X \sim$ 0.7.  

Let us now consider the states belonging to the energy region between 11 and 16 MeV in $^{60}$Ca and let us choose as an illustration the collective peak located at 14.1 MeV. Even if this state still has a neutron dominance, the qualitative prediction for 
the isospin asymmetry of the oscillating system is  lower, $\sim$ 0.6. This qualitative prediction  is indicated by a magenta arrow in the figure. 

Such estimations are coherent with the microscopic predictions obtained using Eq. (\ref{xtot}).   
 The values found for all the excitations under study in this Section are listed in Table \ref{tab3}. 

\begin {table} 
\begin{center}
\begin{tabular}{ccc}
\hline
   Nucleus    &  $E$ (MeV) & $X$  \\
 $^{34}$Si  & 11.07 & 0.73     \\ 
 $^{68}$Ni    &  11.02 & 0.78   \\
 $^{60}$Ca    &    8.8  & 0.84  \\
 $^{60}$Ca    &    14.1  & 0.73  \\
\hline
\end{tabular}
\end{center}
\caption{Values of $X$ (third column) for a given excited state (second column) in the three nuclei under study.  }
\label{tab3}
\end {table} 

We see that, as for the qualitative estimation, the highest asymmetry is found for the 8.8--MeV state in $^{60}$Ca, where $X=$ 0.84. The 8.8, 11.07, and 11.02--MeV states for $^{60}$Ca, $^{34}$Si, and 
$^{68}$Ni, respectively, are characterized by values of $X$ in the range 0.73$\div$0.84 (to be compared with 0.69$\div$0.77 of the qualitative estimation). Finally, the 14.1--MeV state of $^{60}$Ca has a value of $X$ equal to 0.73. This value is the same as the one obtained for $^{34}$Si (whereas, in the previous qualitative estimation, we have found a more significant difference in the isospin asymmetry between the 14.1--MeV state in $^{60}$Ca and the other states in $^{34}$Si, $^{68}$Ni and $^{60}$Ca).  
Note, however, that the 14.1--MeV state in $^{60}$Ca
is strongly mixed (it is only 14\% 1p1h) and, thus, its isospin asymmetry (estimated here taking into account only 1p1h contributions) may be expected to be significantly modified  by the 2p2h contribution. The estimation of $X$ obtained for this state using Eq. (\ref{xtot}) is then less meaningful than for the other states under study which have a more pronounced 1p1h nature.  

\section{Conclusions}
\label{conclu}
We analyzed low--lying isoscalar monopole excitations with a special focus on some selected neutron--rich nuclei, employing the beyond--mean--field SSRPA model based on the Skyrme interaction SGII.
For the results discussed for Ca isotopes, a comparison was shown between RPA and SSRPA spectra. A general trend was found: the excitation modes are predicted both in RPA and in SSRPA. However, the SSRPA excitations are in all cases slightly shifted to lower energies owing to a mixing between 1p1h and 2p2h configurations. This mixing induces a higher fragmentation and renders the nature of these excitations more complex in the SSRPA model. Similar remarks may be done by comparing RPA and SSRPA spectra also for the other nuclei under study in this work and these comparisons were not shown here. 

\begin{widetext}

\begin{figure}
\includegraphics[scale=0.45]{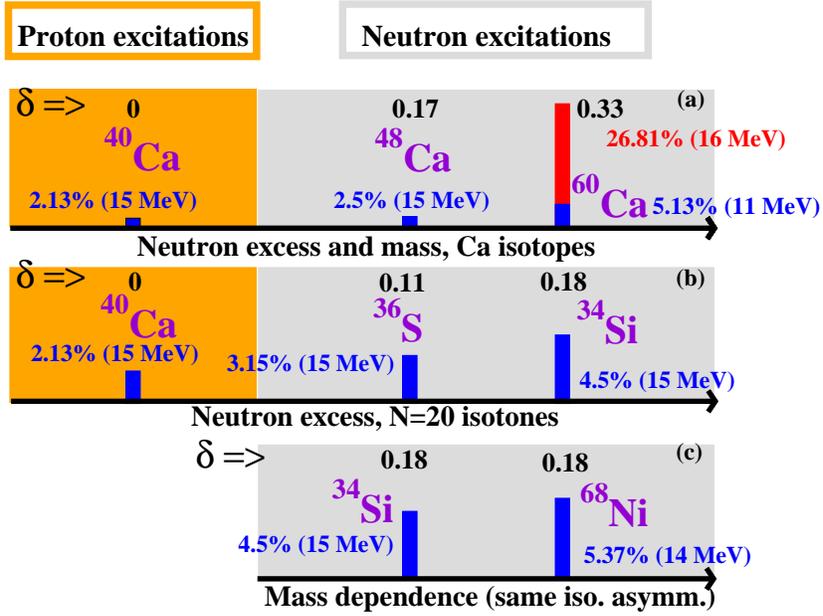}
\caption{Percentages of the EWSR computed up to the energy value written in parentheses for several nuclei. The isospin asymmetry $\delta$ corresponding to each nucleus is indicated. (a) Ca isotopes: evolution of the percentage as a function of the neutron excess and the mass; (b) $N=20$ isotones: evolution of the percentage as a function of the neutron excess; (c) Evolution as a function of the mass for two nuclei with the same isospin asymmetry, $^{34}$Si and $^{68}$Ni.}
\label{summa}
\end{figure}

\end{widetext}

The evolution of the low--lying monopole strength was first studied in Ca isotopes, by increasing the neutron number from $N=20$ in $^{40}$Ca to $N=28$ and $N=40$ in $^{48}$Ca and $^{60}$Ca. A low--lying strength was predicted around 14 MeV in both $^{40}$Ca and $^{48}$Ca. The main peaks are not very collective in both nuclei. The main difference is that  the excitation is driven by protons in $^{40}$Ca whereas, in $^{48}$Ca, it is driven by neutron configurations extending over the entire volume of the nucleus, with one dominant neutron 1p1h configuration. 
The extreme case of $^{60}$Ca was also studied, where the neutron excess is much higher. The soft breathing modes maintain the same nature as in $^{48}$Ca but with a more pronounced neutron nature. Such modes are distributed in two energy regions, the first one being at energies lower than 10 MeV. The collectivity of such excitations is strongly enhanced in $^{60}$Ca compared to $^{48}$Ca. 

A soft monopole mode driven by neutrons was also predicted in the $N=20 $ neutron--rich isotones $^{36}$S and $^{34}$Si. 
To explore the same mass region as $^{60}$Ca but with a smaller isospin asymmetry (comparable to the one of $^{48}$Ca and $^{34}$Si) the nucleus $^{68}$Ni was investigated. The monopole response of this nucleus was already studied in previous analyses both theoretically and experimentally (but without definite conclusions on the existence of a soft monopole mode from the measurement). At variance with Ref. \cite{hamamoto}, which excluded the existence of a soft breathing mode in $^{68}$Ni, we confirm in this work the results of Refs. \cite{khan2011,khan2013} where this mode was predicted. The latter predictions were provided based  on a mean--field model. We use in the present work a model which encompasses beyond--mean--field effects. This enriches the nature of such soft neutron excitations owing to the mixing with 2p2h configurations and leads to a sligth different energy location of the peaks as well as to a higher fragmentation of the strength. 

 The disagreement between the findings of Ref. \cite{hamamoto} and the results of  Refs. \cite{khan2011,khan2013} is maybe related to the fact that the escape width is included in Ref. \cite{hamamoto}, due to the treatment of continuum states. In our case, as in Refs. \cite{khan2011,khan2013}, such a contribution to the total width of an excited mode is not taken into account because continuum is not included. On the other side, in our study, the spreading width is properly accounted for, owing to the coupling with 2p2h configurations (such a contribution to the width is missing in Refs. \cite{hamamoto,khan2011,khan2013}). The difference between our results and those of Ref. \cite{hamamoto} is thus maybe 
related to the fact that we include a spreading width whereas the authors of Ref. \cite{hamamoto} include an escape width. It would be interesting to have a theoretical model taking into account both contributions.

We summarize in Fig. \ref{summa} the evolution of the low--energy contribution to the EWSR for Ca isotopes as a function of the mass and the neutron excess (a), for $N=20$ isotones as a function of the neutron excess (b), and for two nuclei with the same isospin asymmetry but different masses (c). 

Finally, we have discussed a qualitative link between the excitation energies of neutron--driven soft breathing modes and a compressibility modulus introduced for neutron--rich matter. This link also shows that the excitation energies of such predominantly neutron excited states are indeed expected to be lower than the centroids of giant monopole resonances (for the same nucleus) where neutrons and protons participate together and coherently (and where the excitation energies may be related to the compressibility of symmetric nuclear matter). This link between the soft modes and neutron--rich matter may lead to important constraints (coming from measurements of soft modes) on the theoretical models employed for describing neutron--rich matter which may be eventually used in astrophysical applications, for instance for the study of neutron star crusts. 

It is also important to stress that these soft modes should not be included in the estimations of the compressibility modulus of symmetric infinite matter done from measurements: the breathing system is not composed by neutrons and protons oscillating coherently (where the restoring force may be related to the compressibility of symmetric matter) but mainly by neutrons. 

Apart from the extreme case of $^{60}$Ca, measurements for detecting these soft modes are in principle feasible for all the other nuclei studied in this work ($^{34}$Si and $^{68}$Ni are short--lived but measurements are possible). 

As an additional conclusive remark, since we have predicted a proton--driven low--energy strength in the  nucleus $^{40}$Ca (where the neutron excess is equal to zero), we stress that,  
in order to really identify experimentally a low--lying soft monopole mode driven by neutrons in neutron--rich systems, one should conceive and design an experimental setup where the probe is able to evidentiate the dominant neutronic nature. 

\begin{acknowledgments}
This project has received funding from the European Union Horizon 2020 research and innovation program under Grant No. 654002. M.G. acknowledges fruitful discussions with Denis Lacroix.  
\end{acknowledgments}

\end{document}